\documentstyle[12pt]{article}

\def\apj#1{{\em Astrophys. J.,} {\bf #1}}
\def\apjss#1{{\em Astrophys. J. Suppl. Ser.,} {\bf #1}}
\def\aa#1{{\em Astron. \& Astrophys.,} {\bf #1}}
\def\araa#1{{\em Annu. Rev. Astron. \& Astrophys.,} {\bf #1}}

\def\mn#1{{\em Mon. Not. Roy. Astr. Soc.,} {\bf #1}}
\def\kfnt#1#2{{\em Kinematika i Fizika Nebesnykh Tel,} {\bf #1}, {No. #2}}
\def\ass#1{{\em Astrophys. \& Sp. Sci.,} {\bf #1}}

\begin{document}

\begin{center}

\bigskip 

{\Huge {\bf SPH code for dynamical and chemical evolution of disk 
galaxies.}} \\

\bigskip 
\bigskip 

{\Large {\bf Peter Berczik}} \\ 

\bigskip 
\bigskip 

{\Large Main Astronomical Observatory} \\
{\Large Ukrainian National Academy of Sciences} \\
{\Large 252650, Golosiiv, Kiev-022, Ukraine} \\

\bigskip 
\bigskip 

{\Large e-mail: {\tt berczik@mao.kiev.ua}} \\

\bigskip 
\bigskip 

{\Large June 22, 1998} \\

\bigskip 
\bigskip 

\end{center}




\section{Introduction}

\begin{itemize}
   
   \item Galaxy formation in the Universe -- collapse the baryons within 
   DMH potential wells (White \& Rees (78)). Observational support: COBE 
   detection (Bennett et al. (93)).
   
   \item Formation of self -- gravitating inhomogeneities of 
   protogalactic size (Dar (95)). Origin of initial angular momentum 
   (Steinmetz \& Bartelmann (95)).
   
   \item Smoothed Particle Hydrodynamics (SPH) (Monaghan (92)). TREE -- 
   SPH code (Hernquist \& Katz (89)). GRAPE -- SPH code (Steinmetz \& 
   Muller (94, 95)). 

   \item Extension of our N -- body/SPH method (Berczik \& Kolesnik 
   (96), Berczik \& Kravchuk (96)). New "energetic" criteria for SF and 
   more realistic account of returned chemical enriched gas fraction via 
   SNII, SNIa and PN events. 

\end{itemize}


\section{The CD--SPH code}

\begin{itemize}

   \item The SPH code. Continuous hydrodynamic fields in SPH are 
   described by the interpolation functions constructed from the known 
   values of these functions at randomly positioned particles (Monaghan 
   (92)). 
   
   $$
   \rho({\bf r}_i) = \sum_{j=1}^{N} m_j \cdot \frac{1}{2} \cdot
   [W(r_{ij}; h_i) + W(r_{ij}; h_j)].
   $$
   
   The equations of motion for particle $ i $ are:
   
   $$
   \frac{d{\bf r}_i}{dt} = {\bf v}_i.
   $$
   
   $$
   \frac{d{\bf v}_i}{dt} = -\frac{\nabla_i P_i}{\rho_i} +
                    {\bf a}^{vis}_{i} -
                    \nabla_i \Phi_i - 
                    \nabla_i \Phi^{ext}_i.
   $$

   The energy equation in the particle representation has the form:
   
   $$
   \frac{du_i}{dt} = \frac{P_i}{\rho_i} \cdot \nabla_i {\bf v}_i + 
             \frac{\Gamma_i(\rho_i,T_i) - \Lambda_i(\rho_i,T_i)}{\rho_i}.
   $$
   
   The system of equations is closed by adding the equation of state:
   
   $$ 
   P_i = \rho_i \cdot (\gamma - 1) \cdot u_i.
   $$ 

   \item Time integration. To solve the system of equations we use the 
   standard algorithm of leapfrog integrator (Hernquist \& Katz (89)). The 
   integrator has a second order accuracy in the time step $ \Delta t $. 
   To define $ \Delta t $ we use the relation (Hiotelis \& Voglis (91)):
   
   $$
   \Delta t = \min_{i} \{ C_n \cdot 
                \min 
                [\sqrt{\frac{h_i}{\mid {\bf a}_i \mid}},
                \frac{h_i}{\mid {\bf v}_i \mid},
                \frac{h_i}{c_i}] \}.
   $$

   \item The star formation algorithm. We modify the standard SPH star 
   formation algorithm (Katz (92)), taking into account the presence of 
   chaotic motions in the gaseous environment and the time lag between 
   initial development of suitable conditions for star formation and 
   star formation itself (Berczik \& Kravchuk (96)). It states that in 
   the separate "gas" particle the SF can start if the absolute value of 
   the "gas" particle gravitational energy exceeds the sum of its 
   thermal energy and energy of chaotic motions:
   
   $$
   \mid E_i^{gr} \mid > E_i^{th} + E_i^{ch}.
   $$ 
   
   It seems reasonable that the chosen "gas" particle produce stars only 
   if the above condition holds over the time interval exceeding its free 
   -- fall time:
   
   $$ 
   t_{ff} = \sqrt { \frac{3 \cdot \pi}{32 \cdot G \cdot \rho} }.
   $$ 
   
   We check the number of SF acts in selected "gas" particle $ i $. If 
   the number of SF acts becomes greater than $ N^{SF}_{max} = 25 $ we 
   stop any SF activity in these particles. 
   
   We also define which "gas" particles remain cool, i.e. $ t_{cool} < 
   t_{ff} $. These conditions we rewrite in the manner presented in the 
   paper by Navarro \& White (93): $ \rho_i > \rho_{crit} $. Here we use 
   the value of $\rho_{crit} = 0.03$ cm$^{-3}$. 
   
   When the collapsing particle $ i $ has been defined we create the new 
   "star" particle with mass $ m^{star} $ and updated the "gas" 
   particles $ m_i $ using these simple equations: 
   
   $$
   \left\{
   \begin{array}{lll}
   m^{star} = \epsilon \cdot m_i,   \\
                                   \\
   m_i = (1 - \epsilon) \cdot m_i.  \\
   \end{array}
   \right.
   $$
   
   In our Galaxy on the scale of giant molecular clouds the typical 
   values for SF efficiency are in the range $ \epsilon \approx 0.01 
   \div 0.4 $ (Duerr et al. (82), Wilking \& Lada (83)). We define $ 
   \epsilon $ as:
   
   $$ 
   \epsilon = 1 - \frac{E_i^{th} + E_i^{ch}}{\mid E_i^{gr} \mid}.
   $$ 
   
   In the code, we set the absolute maximum value of the mass of such 
   "star" particles $ m^{star}_{max} = 10^7 \; M_\odot $.
   
   At the moment of the birth, the positions and velocities of new 
   "star" particles are set equal to thoes of parent "gas" particles. 
   Subsequently, the "star" particle interact with the rest of "gas" and 
   "star" particles or "dark -- matter" only by gravitation. The 
   gravitational smoothing length for these (Plummer -- like) particles 
   is set equal to $ h_{star} $.
   
   \item The thermal SNII feed -- back. For the thermal budget of the 
   ISM, SNIIs play main role. Following to Katz (92) and Friedli \& Benz 
   (95), we assume that the explosion energy is converted totally to the 
   thermal energy. The total energy released by SNII explosions ($ 
   10^{44} \; J $ per SNII) within "star" particles is calculated at 
   each time step and distributed uniformly between the surrounding 
   (i.e. $ r_{ij} < h_{star} $) "gas" particles (Raiteri et al. (96)). 

   \item The chemical enrichment of gas. In our SF scheme, every new 
   "star" particle represents a separate, gravitationally closed, star 
   formation macro region (like a globular clusters). The "star" 
   particle has its own time of birth $ t_{begSF} $ which is set equal 
   to the moment of the particle formation. After formation of these 
   particles due to SNII, SNIa and PN events, return the chemically 
   enriched gas to surrounding "gas" particles. For description of this 
   process we use the aproximation proposed by Raiteri et al. (96). We 
   concentrate our treatment only on the production of $^{16}$O and 
   $^{56}$Fe, but try to describe the full galactic time evolution of 
   these elements, from the beginning up to present time (i.e. $ 
   t_{evol} \approx 13.0 $ Gyr). 
   
   For example, if the mass of new "star" particle (with metallicity $ Z 
   = 10^{-4} $) is equal to $ 10^4 \; M_\odot $, it produces next 
   numbers of events: $ \Delta N_{\rm SNII} \approx 52.5, \; \Delta 
   N_{\rm PN} \approx 1770, \; \Delta N_{\rm SNIa} \approx 8.48 $ during 
   total time of evolution $ t_{evol} $.

   The total masses (H, He, $^{56}$Fe, $^{16}$O) returned to the 
   surrounding "gas" particles, due to these processes the are (in solar 
   masses): $ \Delta m^{\rm H} \approx 2644, \; \Delta m^{\rm He} 
   \approx 881, \; \Delta m^{\rm Fe} \approx 8.8, \; \Delta m^{\rm O} 
   \approx 120 $.
   
   \item The cold dark matter halo. In the literature we have found 
   some, sometimes contraversial, profiles of Cold Dark Matter Haloes 
   (CDMH) in the galaxies (Burkert (95), Navarro (98)). For resolved 
   structures of CDMH: $ \rho_{halo}(r) \sim r^{-1.4} $ (Moore et al. 
   (97)). The structure of CDMH, as shown in high-resolution N -- body 
   simulations, can be described by: $ \rho_{halo}(r) \sim r^{-1} $ 
   (Navarro et al. (96), Navarro et. al (97)). Finally, in paper by 
   Kravtsov et al. (97) we find that the cores of DM dominated galaxies 
   may have a central profiles: $ \rho_{halo}(r) \sim r^{-0.2} $. 
     
   In our calculations, as a first order aproximation, it is assumed 
   that the model galaxy halo contains the CDMH component with Plummer 
   -- type density profiles (Douphole \& Colin (95)). Therefore for the 
   external force which exercises onto the "gas" and "star" particles by 
   CDMH we can write as:

   $$  
   - \nabla_i \Phi^{ext}_i = - G \cdot \frac{ M_{halo} }
                            { ({\bf r}^2_{i} + b^2_{halo})^{\frac{3}{2}} }
                            \cdot {\bf r}_{i}.
   $$ 
   
\end{itemize}


\section{Results and discussion}

\begin{figure*}[t]

\vspace{13.0cm}
\includegraphics{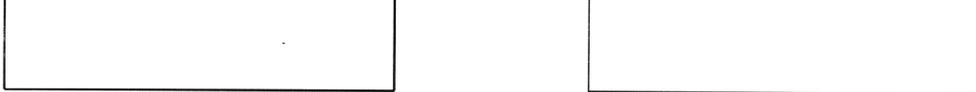}

\caption{The distribution of "star" and "gas" particles in the final step.}
 
\label{fig-xyz}
\end{figure*}

\begin{figure*}[t]

\vspace{10.0cm}
\includegraphics{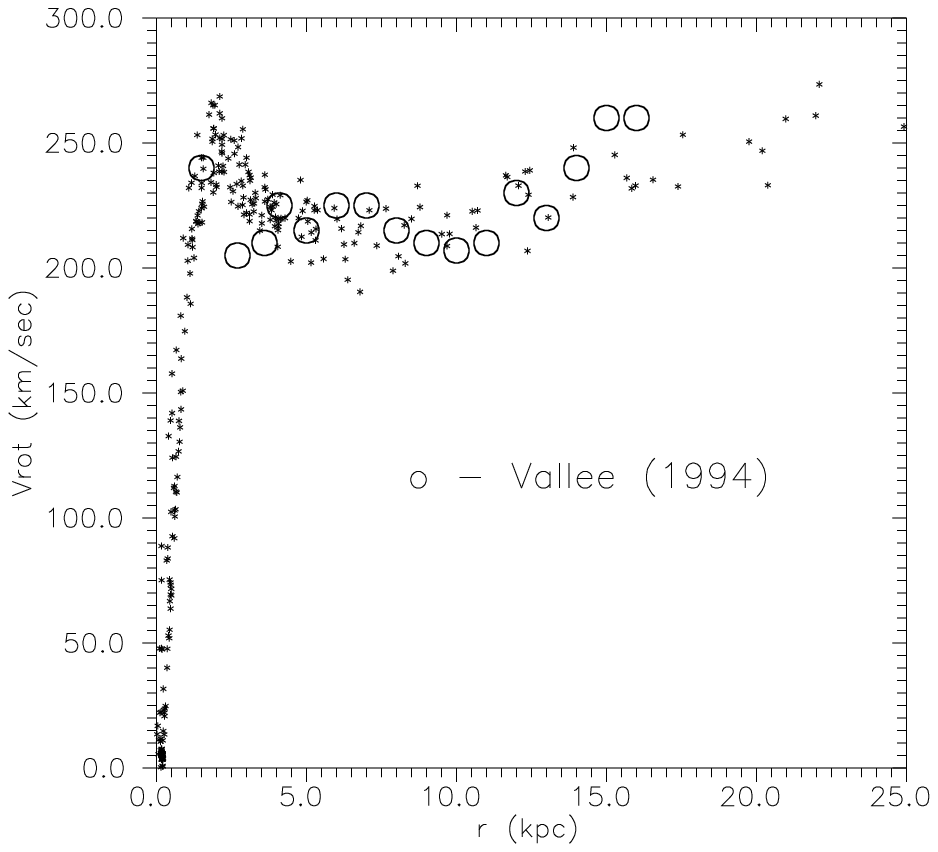}

\caption{The rotational velocity distribution of gas in the final step.}
 
\label{fig-v_rot}
\end{figure*}

\begin{figure*}[t]

\vspace{10.0cm}
\includegraphics{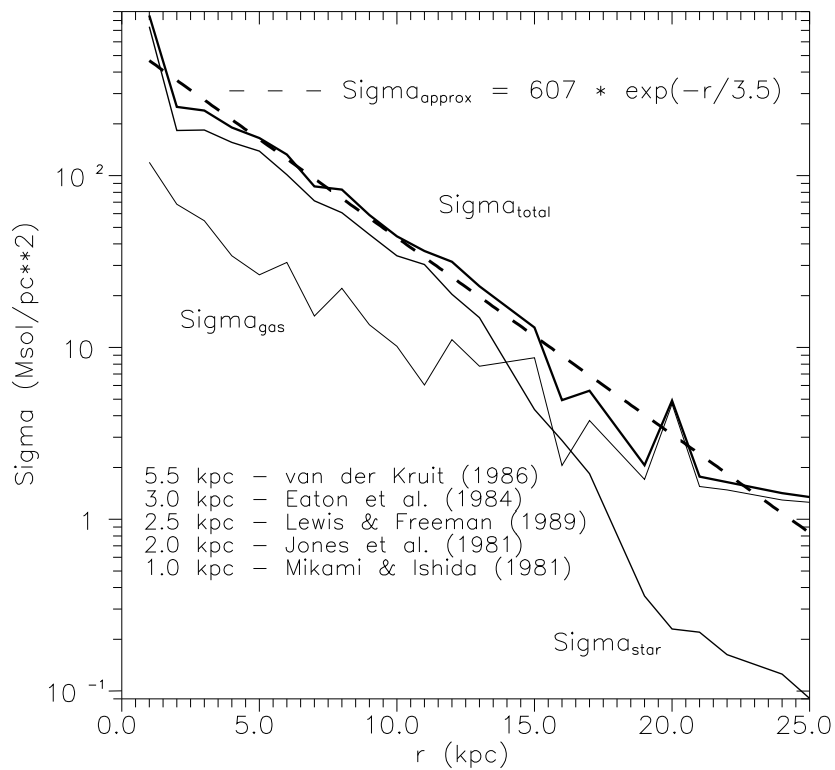}

\caption{The column density distribution in the disks of gas and stars 
in the final step.}
 
\label{fig-sigma}
\end{figure*}

\begin{figure*}[t]

\vspace{10.0cm}
\includegraphics{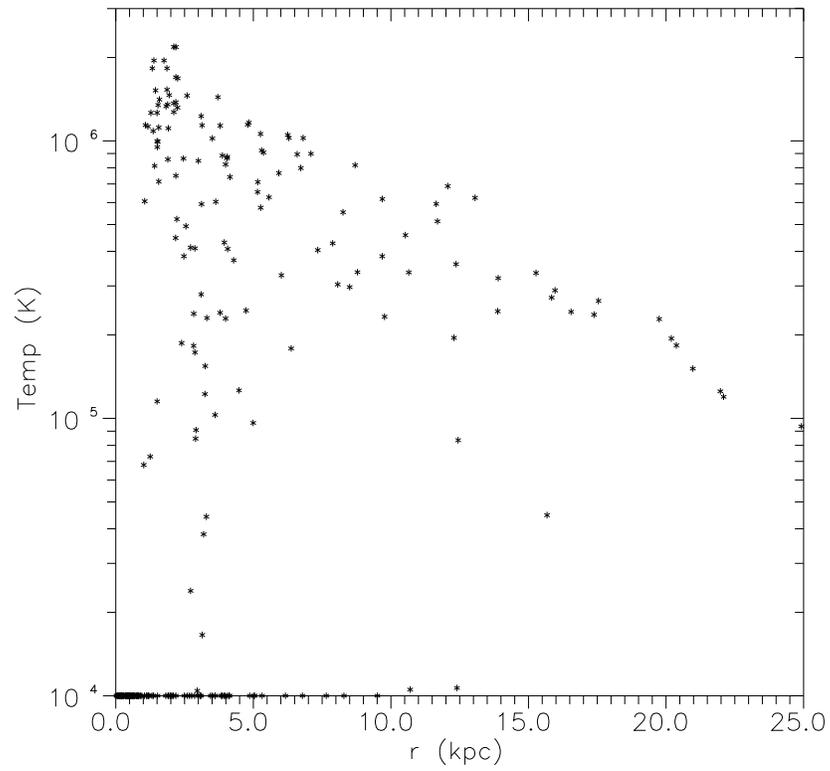}

\caption{The temperature distribution of gas in the final step.}
 
\label{fig-temp}
\end{figure*}

\begin{figure*}[t]

\vspace{10.0cm}
\includegraphics{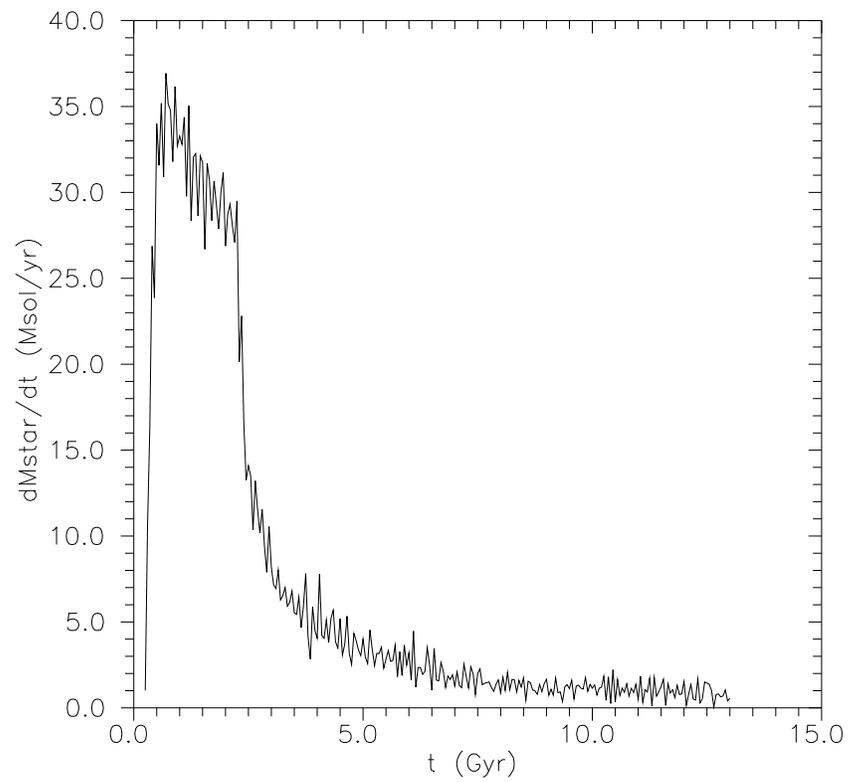}

\caption{The time evolution of the SFR in galaxy.}
 
\label{fig-dm}
\end{figure*}

\begin{figure*}[t]

\vspace{10.0cm}
\includegraphics{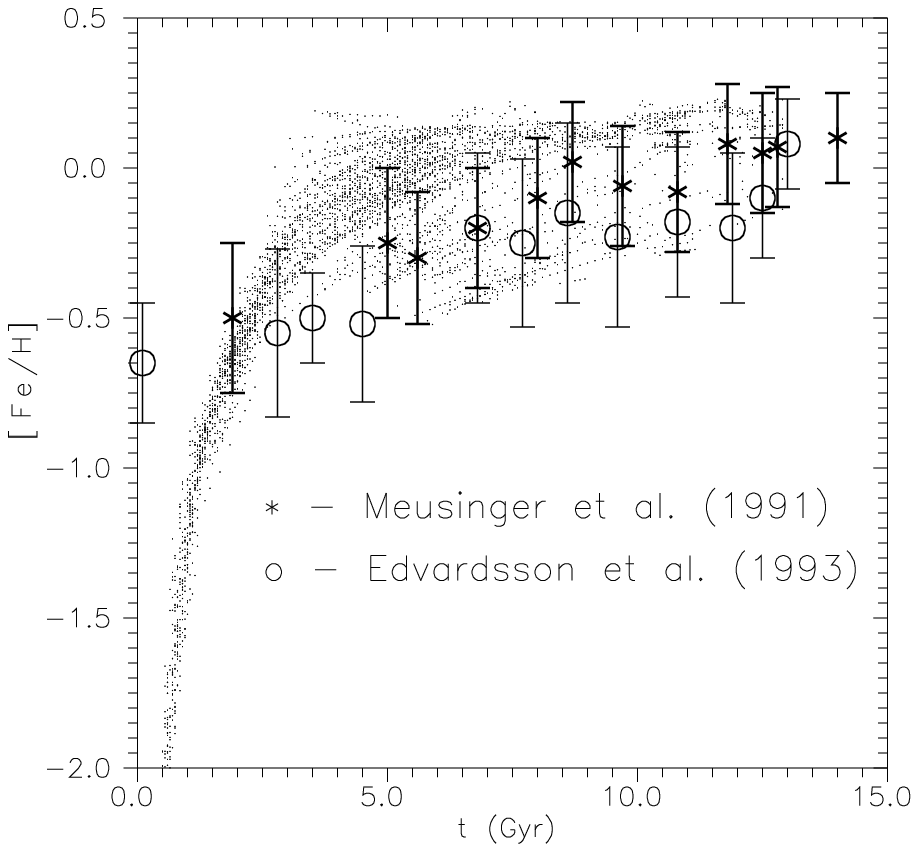}

\caption{The age metallicity realation of the "star" particles in the 
"solar" cylinder ($ 8 $ kpc $ < \; r \; < \; 10 $ kpc).}
 
\label{fig-chem_1}
\end{figure*}

\begin{figure*}[t]

\vspace{10.0cm}
\includegraphics{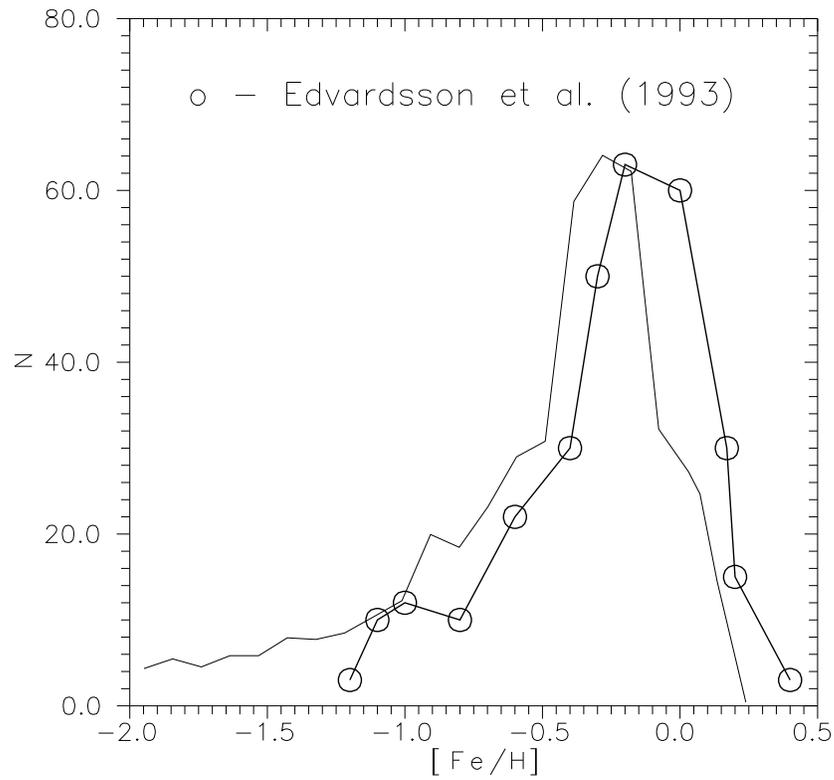}

\caption{The metallicity distribution of the "star" particles in the 
"solar" cylinder ($ 8 $ kpc $ < \; r \; < \; 10 $ kpc).}
 
\label{fig-chem_2}
\end{figure*}

\begin{figure*}[t]

\vspace{10.0cm}
\includegraphics{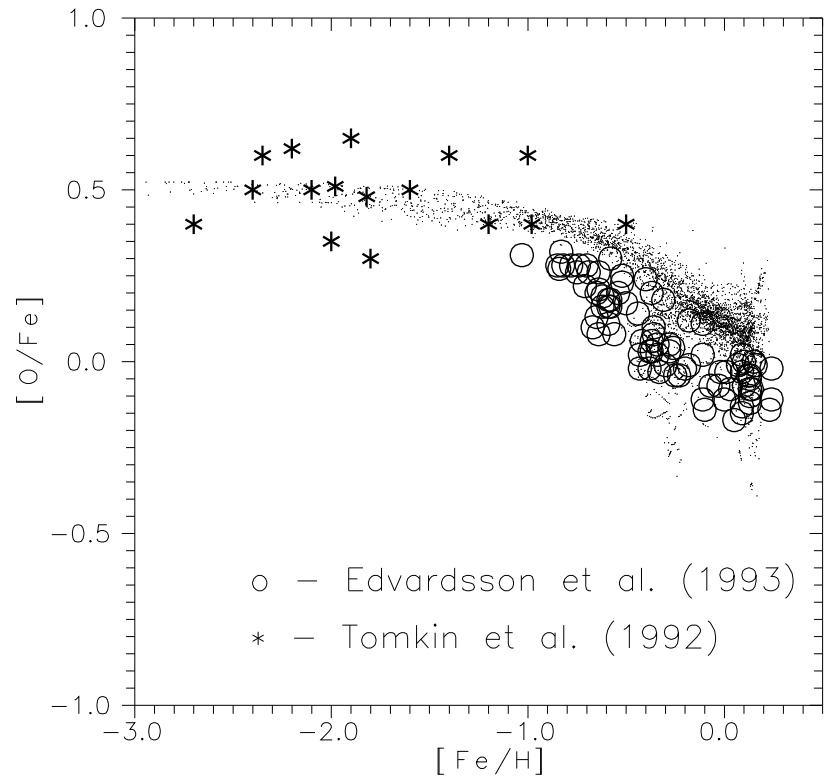}

\caption{The [O/Fe] vs. [Fe/H] distribution of the "star" particles in 
the "solar" cylinder ($ 8 $ kpc $ < \; r \; < \; 10 $ kpc).}
 
\label{fig-chem_3}
\end{figure*}

\begin{figure*}[t]

\vspace{10.0cm}
\includegraphics{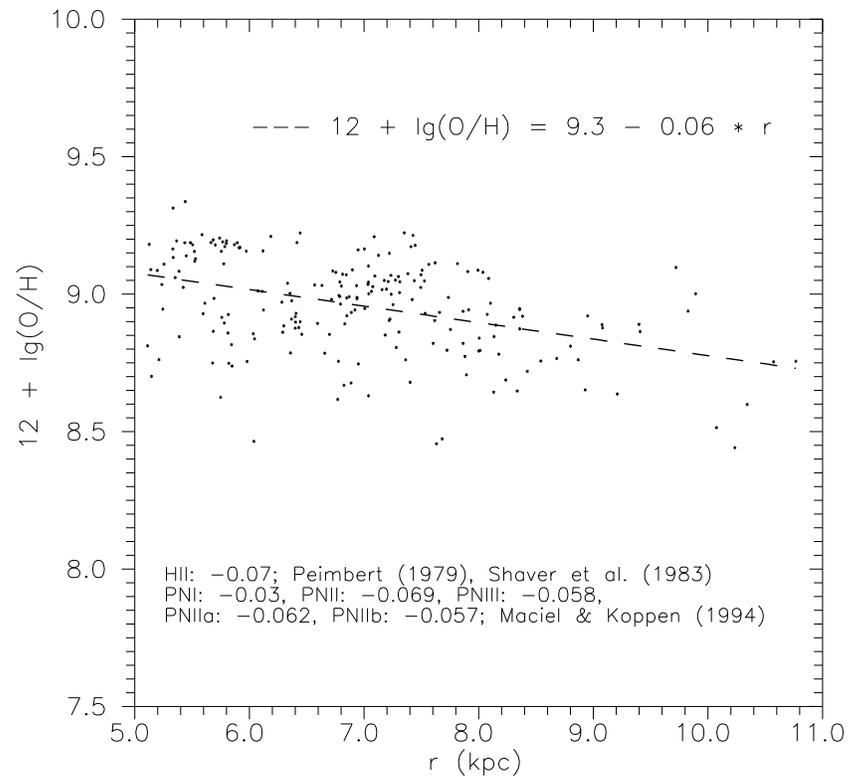}

\caption{The [O/H] radial distribution.}
 
\label{fig-grad-z}
\end{figure*}

\begin{itemize}

   \item Initial conditions. The SPH calculations were carried out for $ 
   N_{gas} = 2109 $ "gas" particles. According to Navarro \& White (93) 
   and Raiteri et al. (96), such number seems to be quite enough to 
   provide qualitatively correct description of the system behaviour. 
   Even such small number of "gas" particles produces a $ N_{star} = 
   31631 $ "star" particles at the end of calculation. 

   The value of the smoothing length $ h_i $ was chosen requiring that 
   each "gas" particle had $ N_B = 21 $ neighbours within $ 2 \cdot h_i $. 
   Minimal $ h_{min} $ was set equal to $ 1 $ kpc. For "star" particles 
   we use the fixed gravitational smoothing length $ h_{star} = 1 $ kpc. 
   
   As initial model (relevant for CDM -- scenario) we took constant -- 
   density homogeneous triaxial configuration of gas ($ M_{gas} = 
   10^{11} \; M_\odot $) within the dark matter halo ($ M_{halo} = 
   10^{12} \; M_\odot $). We set $ A = 100 $ kpc, $ B = 75 $ kpc  
   and $ C = 50 $ kpc for semiaxes of system . We set the smoothing 
   parameter of CDMH: $ b_{halo} = 25 $ kpc. The gas component was 
   assumed to be initially cold, $ T_0 = 10^{4} $ K. 
   
   The gas was assumed to be involved into the Hubble flow ($ H_{0} = 65 
   $ km/s/Mpc, $ {\Omega}_{0} = 1) $ and into the solid -- body rotation 
   around $ z $ -- axis. We added the small random components of 
   velocities ($ \Delta \mid {\bf v} \mid = 10 $ km/s) to account for 
   the chaotic motions of fragments. 
   
   The spin parameter in our simulation is $ \lambda \approx 0.08 $. 
   This parameters is defimed in Peebles (69) as:

   $$
   \lambda = \frac{\mid {\bf L}_0 \mid \cdot \sqrt{\mid E_0^{gr} \mid}}
              {G \cdot (M_{gas}+M_{dm})^{5/2}}.
   $$

   If the angular momentum is acquired through the tidal torque of the 
   surrounding matter, the standard spin parameter does not exceed $ 
   \lambda \approx 0.11 $ (Steinmetz \& Bartelmann (95), $ \lambda 
   \approx 0.07^{+0.04}_{-0.05} $).
   
   \item Conclusion. This simple model provides good, self -- 
   consistent picture of the process of galaxy formation. The dynamical 
   and chemical evolution of modelled disk -- like galaxy is coincident
   with the results of observations for our own Galaxy. Some basic 
   distributions of gas and star parameters are given in figures:
   
   \begin{itemize}
   
   \item Fig.~\ref{fig-xyz}. The distribution of "star" and "gas" 
   particles in the final step.
   
   \item Fig.~\ref{fig-v_rot}. $ V_{rot}(r) $. The rotational velocity 
   distribution of gas in the final step.
   
   \item Fig.~\ref{fig-sigma}. $ \sigma_{*}(r) $,~~~$ \sigma_{gas}(r) $. 
   The column density distribution in the disks of gas and stars in the 
   final step.
   
   \item Fig.~\ref{fig-temp}. $ T(r) $. The temperature distribution of 
   gas in the final step.
   
   \item Fig.~\ref{fig-dm} $. SFR(t) $. The time evolution of the SFR in 
   galaxy.
   
   \item Fig.~\ref{fig-chem_1}. [Fe/H]$(t)$. The age metallicity 
   relation of the "star" particles in the "solar" cylinder ($ 8 $ kpc $ 
   \; < \; r \; < \; 10 $ kpc).
   
   \item Fig.~\ref{fig-chem_2}. $ N_{*}($[Fe/H]$) $.The metallicity 
   distribution of the "star" particles in the "solar" cylinder ($ 8 $ 
   kpc $ < \; r \; < \; 10 $ kpc).
   
   \item Fig.~\ref{fig-chem_3}. [O/Fe]([Fe/H]). The [O/Fe] vs. 
   [Fe/H] distribution of the "star" particles in the "solar" cylinder 
   ($ 8 $ kpc $ < \; r \; < \; 10 $ kpc).
   
   \item Fig.~\ref{fig-grad-z}. [O/H]$(r) $. The [O/H] radial 
   distribution.
   
   \end{itemize}
   
\end{itemize}


\bigskip

{\Large \centerline{\bf Acknowledgements:}}

\bigskip
     
{\Large The author are grateful to S.G. Kravchuk, L.S. Pilyugin, Yu.I. 
Izotov and S.A. Silich for fruitful discussions during the process of 
preparing this work and would like to acknowledge the American 
Astronomical Society for International Small Research Grant (1998).}



\eject

\eject

\eject

\eject

\end{document}